\documentclass[onecolumn,showpacs]{revtex4}

\usepackage{graphicx}
\usepackage{dcolumn}
\usepackage{bm}

\input epsf

\begin{document}

\title{\Large ~Brane world solutions of perfect fluid in the background
of a bulk containing dust or cosmological constant}

\author{\bf~Tanwi~Bandyopadhyay$^1$,~Subenoy~Chakraborty$^1$\footnote{subenoy@iucaa.ernet.in}
~and~Asit~Banerjee$^2$}

\affiliation{$^1$Department of Mathematics,~Jadavpur
University,~Kolkata-32, India.\\ $^2$Department of
Physics,~Jadavpur University,~Kolkata-32, India.}

\date{\today}

\begin{abstract}
The paper presents some solutions to the five dimensional Einstein
equations due to a perfect fluid on the~brane with pure dust
filling the entire bulk in one case and a cosmological constant
(or vacuum) in the bulk for the second case. In the first case,
there is a linear relationship between isotropic pressure, energy
density and the~brane tension, while in the second case, the
perfect fluid is assumed to be in the form of~chaplygin gas.
Cosmological solutions are found both for~brane and bulk
scenarios and some interesting features are obtained for
the~chaplygin gas on the~brane which are distinctly different
from the standard cosmology in four dimensions.
\end{abstract}


\maketitle

\section{\normalsize\bf{Introduction}}

Recently there has been a proliferation of~brane world models,
where the standard model matter is confined in a four dimensional
space time said to be a singular~hypersurface or $3$-brane
embedded in a $(4+d)$-dimensional space time. This
$(4+d)$-dimensional space time is said to be the bulk. Though the
matter is all confined to a $3$-brane, the gravitational field
can propagate in the higher dimensional bulk. The $d$ extra
dimensions, however, need not be small or even compact in this
theory. In fact, Randall and Sundrum [1] have shown that for
$d=1$ the fifth dimension may be infinite. The effective
equations for gravity in four dimensions were obtained by
Shiromizu et al [2] using Israel's boundary conditions and
$Z_{2}$-symmetry for the bulk space time, in which the~brane is
embedded. It has been demonstrated by Bin\'{e}truy et al [3,4]
that generically the equations governing the cosmological
evolution of the~brane will be different from those corresponding
to the analogous Friedmann equations of standard cosmology.
Essentially the difference lies in the fact that the energy
density of the~brane appears in quadratic form in the~brane
equations, whereas it is linear in the standard cosmology.\\

In the present paper our motivation is to solve the five
dimensional Einstein equations for a perfect fluid on the~brane
while the matter in the bulk is either in the form of pure dust
or a cosmological constant (or vacuum). It is well known that in
the standard cosmology, a special kind of fluid called~chaplygin
gas [5 and references therein] has a very interesting property
that it can drive the standard $4D$ universe from its initial
dust dominated stage to a final $\Lambda$ dominated stage at the
end, so that at the late stage, one can obtain an accelerating
world model consistent with the recent Supernova type Ia
observations [6]. Measurements of luminosity-redshift relation,
obtained from observations suggest that most of the energy in the
universe happens to be dark. Different kinds of the dark energy
viewed from different angles have so far been proposed [7] to
explain the late time acceleration of the universe. Recently,
Ponce de Leon [8-10] studied the acceleration of the $4D$
universe as a consequence of the time evolution of the vacuum
energy or the ~brane tension in the context of the~brane world models.\\

In our paper, section II presents the derivation of~brane
equations starting from the Einstein equations in the bulk with a
general form of matter there. Then in section III, assuming pure
dust in the bulk, Einstein equations in the~brane show that only
a kind of~barotropic equation of state
$p_{b}=-\frac{2}{3}\rho_{b}$ holds between the effective
isotropic pressure $p_{b}$ and effective energy density
$\rho_{b}$ (i.e, a linear relation among the actual pressure and
energy density in the~brane and the~brane tension). Cosmological
solutions are obtained for both bulk and~brane scenario. Section
IV deals with perfect fluid in the form of~chaplygin gas in the
~brane model. When the bulk is vacuum, then the dynamics shows
initial~deceleration of the~brane with subsequent acceleration
going over to the $\Lambda$CDM model. On the other hand, when the
bulk has negative $5D$ cosmological constant, i.e, an anti
de-Sitter bulk, there occur two different cases. Depending on the
parameters in the field equations, the~brane may either transit
from a decelerating stage to an accelerating stage in course of
time or it may simply exhibit a~recollapsing behaviour. The paper
ends with the concluding remarks.\\

\section{\normalsize\bf{Basic Equations}}

We shall assume that the geometry of the five dimensional bulk is
characterized by the space time metric of the form

\begin{equation}
ds^{2}=\bar{g}_{\mu\nu}dx^{\mu}dx^{\nu}=g_{ij}dx^{i}dx^{j}+b^{2}dy^{2}
\end{equation}

with $y$ as the fifth coordinate. The~hypersurface $y=0$ is
identified as the world volume of the~brane that forms our
universe. As usual, for simplicity, the explicit form of the
metric (1) is taken to be

\begin{equation}
ds^{2}=-n^{2}(t,y)dt^{2}+a^{2}(t,y)\delta_{ij}dx^{i}dx^{j}+b^{2}(t,y)dy^{2}
\end{equation}

where for simplicity we choose the usual spatial section of
the~brane to be flat.\\

Now for cosmological solutions, we shall have to specify the
matter explicitly. In fact, we shall classify the matter as of
two distinct parts namely\\

i)~matter confined to our~brane universe and\\
ii)~matter distributed over the $5D$ bulk.\\

Hence, the energy momentum tensor can be decomposed into

\begin{equation}
{\bar{T}_{\mu}}^{\nu}={{\tilde{T}_{\mu}}^{\nu}}~_{\text~bulk}
+{{T_{\mu}}^{\nu}}_{\text~brane}
\end{equation}

where the explicit form of the matter is [11,~3,~4]

\begin{equation}
{{\tilde{T}_{\mu}}^{\nu}}~_{\text~bulk}=diag(-\rho_{B},p_{B},
p_{B},p_{B},p_{5})
\end{equation}

and

\begin{equation}
{{T_{\mu}}^{\nu}}_{\text~brane}=\frac{\delta(y)}{b}
diag(-\rho_{b},p_{b},p_{b},p_{b},0)
\end{equation}

In general, energy-momentum tensor in the bulk depends on $t$ and
$y$ (i.e,~$\rho_{B}=\rho_{B}(t,y)$) but to recover a homogeneous
cosmology in the~brane, matter field in the~brane is assumed to
be the function of time alone. Also the~brane is assumed to be
infinitely thin. Now the explicit form of the five dimensional
Einstein equations

\begin{equation}
{\bar{G}_{\mu}}^{\nu}=\kappa^{2}{\bar{T}_{\mu}}^{\nu}
\end{equation}

for the above metric (2) having matter distribution in the form
(3) are [11,~3]

\begin{equation}
\frac{3}{n^{2}}\left[\frac{\dot{a}}{a}\left(\frac{\dot{a}}{a}
+\frac{\dot{b}}{b}\right)-\frac{n^{2}}{b^{2}}\left\{\frac{a''}{a}
+\frac{a'}{a}\left(\frac{a'}{a}-\frac{b'}{b}\right)\right\}\right]
=\kappa^{2}\left(\rho_{B}+\frac{\delta(y)}{b}\rho_{b}\right)
\end{equation}

\begin{eqnarray*}
\frac{1}{b^{2}}\left[\frac{a'}{a}\left(\frac{a'}{a}+2\frac{n'}{n}\right)
-\frac{b'}{b}\left(\frac{n'}{n}+2\frac{a'}{a}\right)+2\frac{a''}{a}
+\frac{n''}{n}\right]
\end{eqnarray*}
\begin{equation}
+\frac{1}{n^{2}}\left[\frac{\dot{a}}{a}\left(-\frac{\dot{a}}{a}
+2\frac{\dot{n}}{n}\right)-2\frac{\ddot{a}}{a}+\frac{\dot{b}}{b}
\left(-2\frac{\dot{a}}{a}+\frac{\dot{n}}{n}\right)-\frac{\ddot{b}}{b}\right]
=\kappa^{2}\left(p_{B}+\frac{\delta(y)}{b}p_{b}\right)
\end{equation}

\begin{equation}
3\left(\frac{n'}{n}\frac{\dot{a}}{a}+\frac{a'}{a}\frac{\dot{b}}{b}
-\frac{\dot{a}'}{a}\right)=0
\end{equation}

and

\begin{equation}
\frac{3}{b^{2}}\left[\frac{a'}{a}\left(\frac{a'}{a}+\frac{n'}{n}\right)
-\frac{b^{2}}{n^{2}}\left\{\frac{\dot{a}}{a}\left(\frac{\dot{a}}{a}
-\frac{\dot{n}}{n}\right) +\frac{\ddot{a}}{a}\right\}\right]
=\kappa^{2}p_{5}
\end{equation}

where an~overdot stands for differentiation with respect to time
$t$ and a prime indicates derivative with respect to the fifth
coordinate $y$.\\

For~brane cosmology, we need solution of the Einstein's equations
(7)-(10) as $y\rightarrow0$. Now, for a well defined geometry,
the metric tensor should be continuous across the~brane (i.e,
$y=0$) while its derivative with respect to $y$ may
be~discontinuous at the ~hypersurface $y=0$. As a result, Dirac
delta function will appear in the second derivatives of the
metric coefficients with respect to $y$ namely [3]

\begin{equation}
a''=(a'')+[a']~\delta(y)
\end{equation}

Here $(a'')$ is the non-distributional part of the second
derivative i.e, the usual second order derivative and $[a']$ is
the measure of discontinuity of first derivative across $y=0$
(with $\delta(y)$, the Dirac delta function) as

\begin{equation}
[a']=a'(0+)-a'(0-)
\end{equation}

If we now proceed to the limit as $y\rightarrow0$ and match the
Dirac delta function in both sides of the field equations (7) and
(8) we obtain

\begin{equation}
\frac{[a']}{a_{0}b_{0}}=-\frac{\kappa^{2}}{3}\rho_{b}
\end{equation}

\begin{equation}
\frac{[n']}{n_{0}b_{0}}=\frac{\kappa^{2}}{3}~(3p_{b}+2\rho_{b})
\end{equation}

where subscript $'0'$ in the scale factors indicate their values
on the~brane.\\

One may note that the jump of the field equation (9) using
equations (13) and (14) results in the energy conservation
relation on the~brane namely

\begin{equation}
\dot{\rho_{b}}+3~(\rho_{b}+p_{b})\frac{\dot{a_{0}}}{a_{0}}=0
\end{equation}

Further, if we take the average value of the field equation (10)
for $y\rightarrow0^{+}$ and $y\rightarrow0^{-}$ and impose the
$Z_{2}$-symmetry we get [3]

\begin{equation}
\frac{\ddot{a_{0}}}{a_{0}}+\frac{\dot{a_{0}}^{2}}{a_{0}^{2}}
=-\frac{\kappa^{4}}{36}~\rho_{b}(\rho_{b}+3p_{b})
-\frac{\kappa^{2}p_{5}}{3b_{0}^{2}}
\end{equation}

Here we have chosen $n_{0}=1$, which is allowed by a suitable
time transformation and hence $t$ is now the usual cosmic time on
the~brane. This equation can be termed as generalized Friedmann
type equation on the~brane. The basic difference with the
corresponding equation in standard cosmology is that, here the
square of the Hubble parameter depends~quadratically on the~brane
energy density, in contrast with the usual linear dependence for
Friedmann universe.\\

Moreover, by introducing a function $\xi(t,y)$ defined by

\begin{equation}
\xi(t,y)=\frac{(a'a)^{2}}{b^{2}}-\frac{(\dot{a}a)^{2}}{n^{2}}
\end{equation}

the five dimensional field equations (7) and (10) can be written
in compact form as (for $y\neq0$)

\begin{equation}
\xi'=-\frac{2}{3}~a'a^{3}\kappa^{2}\rho_{B}
\end{equation}
\begin{equation}
\dot{\xi}=~\frac{2}{3}~\dot{a}a^{3}\kappa^{2}p_{5}
\end{equation}

Now equating the time derivative of equation (18) with the
y-derivative of equation (19) we have

\begin{equation}
a'\dot{\rho_{B}}+\dot{a}p_{5}'+\left(\rho_{B}+p_{5}\right)
\left(\dot{a}'+3\frac{\dot{a}a'}{a}\right)=0
\end{equation}

As we have assumed $\bar{T}_{05}=0$ indicating that there is no
flow of energy along the extra dimension, the Bianchi identity
$\nabla_{\mu}\bar{G}^{\mu0}=0$ results

\begin{equation}
\dot{\rho_{B}}+3\frac{\dot{a}}{a}(\rho_{B}+p_{B})+\frac{\dot{b}}{b}
(\rho_{B}+p_{5})=0
\end{equation}

Also taking average value of the equation(17) for
$y\rightarrow0^{+}$ and $y\rightarrow0^{-}$, imposing the
$Z_{2}$-symmetry and using the junction condition (13) we obtain
the generalized Friedmann equation

\begin{equation}
\frac{\dot{a_{0}^{2}}}{a_{0}^{2}}=\frac{\kappa^{4}}{36}\rho_{b}^{2}
-\frac{\xi_{0}(t)}{a_{0}^{4}}
\end{equation}

where $\xi_{0}$ is the value of $\xi$ on the~brane.\\

Therefore, the cosmological evolution on the bulk is completely
characterized by the Einstein field equations (7)-(10) (also
(18), (19)) and by the conservation equations (20) and (21). On
the other hand,~brane cosmology will be determined by equations
(16) and (22), junction conditions (13) and (14) and by
the matter conservation on the~brane (15).\\

\section{\normalsize\bf{~Brane cosmology for dust filled bulk}}

If the bulk matter is assumed to be in the form of pure dust i.e,
$p_{B}=p_{5}=0$, then integrating the conservation equation (21)
we get

\begin{equation}
\rho_{B}(t,y)=\rho_{1}(y)/a^{3}b
\end{equation}

where $\rho_{1}(y)$ is function of the coordinate $y$ alone.\\
Also comparing the conservation equations (20) and (21) and using
the field equation (9) we have

\begin{equation}
n'=0~\text{(so that $n$ may be assumed to be unity without the
loss of generality)~~~~and}~~~a'=\alpha(y)~b
\end{equation}

with $\alpha$, an arbitrary function of $y$.\\

As $p_{5}=0$, the equation (19) yields $\xi=\beta(y)$, a function
of $y$ alone which again in view of (18) satisfies the relation

\begin{equation}
\beta'(y)=-\frac{\kappa^{2}\rho_{B}}{6}(a^{4})'
\end{equation}

Now using (24) in (17), we obtain the differential equation for
$a$ in the form\\

~~~~~~~~~~~~~~~~~~~~~~~~~~~~~~~~~~~~~~~~~~~~~~~~~~~~~~~
$a^{2}\dot{a}^{2}=\alpha^{2}a^{2}-\beta$\\

which on integration gives the solution for the scale factor
on~brane in the form

\begin{equation}
a^{2}(t,y)=\frac{1}{\alpha^{2}(y)}\left[\beta(y)+\left\{\alpha^{2}(y)t
+t_{0}(y)\right\}^{2}\right]
\end{equation}

with $t_{0}(y)$, an arbitrary integration function.\\

In order to obtain cosmological solutions on the~brane, we now
consider the generalized Friedmann equation (22) written earlier

\begin{equation}
\frac{\dot{a_{0}}^{2}}{a_{0}^{2}}=\frac{\kappa^{4}}{36}\rho_{b}^{2}
-\frac{\beta_{0}}{a_{0}^{4}}
\end{equation}

Further, combining (24) with the junction condition (14) we get
for the effective equation of state for the~brane matter

\begin{equation}
p_{b}=-\frac{2}{3}~\rho_{b}
\end{equation}

Equivalently, as

\begin{equation}
p_{b}=p-\lambda~~~\text{and}~~~\rho_{b}=\rho+\lambda
\end{equation}

with $\lambda$ representing the~brane tension, the above equation
of state yields the following linear relation

\begin{equation}
p=-\frac{2}{3}\rho+\frac{1}{3}\lambda
\end{equation}

Thus for dust bulk, the fluid in the~brane must satisfy the above
linear relation between isotropic pressure, energy density
and~brane tension. Using now the above equation of state (28)
into the~brane conservation equation (15) we get

\begin{equation}
\rho_{b}=\rho_{0}/a_{0}~~,~~~~~~\text{$\rho_{0}$ here is simply a
constant.}
\end{equation}

Now since $a'=\alpha(y)b$~ as is evident from the equation (24),
the~brane must satisfy the following relation \\

~~~~~~~~~~~~~~~~~~~~~~~~~~~~~~~~~~~~~~~~~~~~~~~~~~~$[a']=2b_{0}\alpha_{0}$\\

and hence using equations (13) and (31), the arbitrary constants
$\rho_{0}$ and $\alpha_{0}$ above are shown to be related as

\begin{equation}
\alpha_{0}=-\kappa^{2}\rho_{0}/6
\end{equation}

Further, for the dust model in the bulk and the equation of state
(28) in the~brane, the Friedmann type equation (16) simplifies to

\begin{equation}
2a_{0}\ddot{a_{0}}+\dot{a_{0}}^{2}=\mu~,~~~~\text{where}~~~
\mu=\frac{\kappa^{4}\rho_{0}^{2}}{18}
\end{equation}

which has the first integral

\begin{equation}
\dot{a_{0}}^{2}=\frac{\mu}{2}+\frac{\mu_{0}}{a_{0}^{2}}
\end{equation}

where $\mu_{0}$ is an integration constant. This is fully
consistent with the equation (27) provided
$\mu_{0}=-\beta_{0}$.\\

Now further integration of the equation (34) yields the explicit
expression for the~brane scale factor $a_{0}$ as

\begin{equation}
a_{0}^{2}=2\mu~(t_{1}\pm t)^{2}-\frac{2\mu_{0}}{\mu}
\end{equation}

Here since $\ddot{a_{0}}=-\frac{\mu_{0}}{a_{0}^{3}}$, we always
get either an accelerating model or a decelerating model of
the~brane universe depending on the sign of the constant
$\mu_{0}$, but there can not be any midway transition from the
decelerating phase to the accelerating one (or the reverse). The
choice of sign in front of $t$ in equation (35) leads to an
expanding (+ve sign) or a contracting (-ve sign)~brane world.\\

\section{\normalsize\bf{~Chaplygin gas~Brane model}}

We consider here a modified~chaplygin gas [12], which satisfies an
equation of state in the following form

\begin{equation}
p_{b}=A\rho_{b}-\frac{B}{{\rho_{b}}^{\alpha}}
\end{equation}

where $A$ and $B(>0)$ are constants and $0<\alpha\leq1$.\\

So in this case from the energy conservation equation (15), the
expression for matter density is given by [12]

\begin{equation}
\rho_{b}=\left[\frac{1}{1+A}\left\{\frac{\rho_{2}}{a_{0}^{3(1+A)(1+\alpha)}}
+B\right\}\right]^\frac{1}{1+\alpha}
\end{equation}

with $\rho_{2}$, an arbitrary integration constant.\\

{\bf Case-I:~~Vacuum Bulk}\\

As there is no matter in the bulk we have
${{\tilde{T}_{\mu}}^{\nu}}~_{\text~bulk}\equiv0$ and the field
equations (18) and (19) show that $\xi$ must be a constant (say,
$C$). Hence we have the generalized Friedmann equation (see eq.
(22))

\begin{equation}
\frac{\dot{a_{0}^{2}}}{a_{0}^{2}}=\frac{\kappa^{2}}{36}~\rho_{b}^{2}
+\frac{C}{a_{0}^{4}}
\end{equation}

Also the generalized Friedmann type equation (16) now simplifies
to

\begin{equation}
\frac{\ddot{a_{0}}}{a_{0}}+\frac{\dot{a_{0}^{2}}}{a_{0}^{2}}
=-\frac{\kappa^{4}}{36}~\rho_{b}(\rho_{b}+3p_{b})
\end{equation}

Further, our calculation will be much simplified if we choose
$\alpha=1$ in the equation of state (36) and hence from (37) we
get the differential equation for the scale factor $a_{0}$ in the
form

\begin{equation}
\ddot{a_{0}}=-\frac{\kappa^{4}}{36}\frac{(3A+2)}{(A+1)}
\frac{\rho_{0}}{a_{0}^{(6A+5)}}+\frac{\kappa^{4}Ba_{0}}{36(1+A)}
-\frac{C}{a_{0}^{3}}
\end{equation}

Note that as $a_{0}\rightarrow0$ (early epoch)
$\rho_{b}\rightarrow\infty,~\ddot{a_{0}}\rightarrow-\infty$ while
as $a_{0}\rightarrow\infty,
~\rho_{b}\rightarrow\left(\frac{B}{1+A}\right)^{1/2}$ which in
view of (36) leads to
$p_{b}\rightarrow-\left(\frac{B}{1+A}\right)^{1/2}$ and
$\ddot{a_{0}}\rightarrow\infty$. This shows that asymptotically
the equation of state reduces to $p_{b}=-\rho_{b}$, which points
to a $\Lambda$CDM model in the corresponding FRW universe of
standard cosmology. But for some special cases of the~brane world
models (shown in what follows) the situation is different, since
here $\ddot{a_{0}}$ may remain negative throughout even if the
scale factor $a_{0}$ increases indefinitely. Such behaviour is
not consistent in the four dimensional Friedmann model containing
~chaplygin gas, which always asymptotically goes over to the
$\Lambda$CDM model with the increase of the scale factor.\\

We now proceed to solve equation (40) for $A=\frac{1}{3}$ (for
which, solution in closed form is possible). The first integral of
equation (40) can be written in an integral form as

\begin{equation}
\frac{1}{4}\int\frac{dz}{\sqrt{bz^{2}+Cz+d}}=\pm~(t-t_{0})
\end{equation}

with
$z=a_{0}^{4},~b=\frac{\kappa^{4}B}{48},~d=\frac{\kappa^{4}\rho_{0}}{48}$,
solutions of which are given by

\begin{equation}
a_{0}^{4}=\frac{\sqrt{4bd-C^{2}}}{2b}~Sinh\left[\pm4\sqrt{b}~(t-t_{0})\right],
~~~~~~~~~~~~~~~~(\text{when $4bd-C^{2}>0$})
\end{equation}

\begin{equation}
a_{0}^{4}=\frac{\sqrt{C^{2}-4bd}}{2b}~Cosh\left[\pm4\sqrt{b}~(t-t_{0})\right],
~~~~~~~~~~~~~~~~(\text{when $4bd-C^{2}<0$})
\end{equation}

and

\begin{equation}
a_{0}^{4}=\frac{1}{2b}\left[e^{\pm4\sqrt{b}~(t-t_{0})}\right]-\frac{C}{2b},
~~~~~~~~~~~~~~~~(\text{when $4bd-C^{2}=0$})
\end{equation}

Graphically, it can be seen that for equation (42), the scale
factor expands exponentially whereas for equation (43), it
decreases to a minimum before further expansion.\\

\begin{figure}
\includegraphics[height=2in]{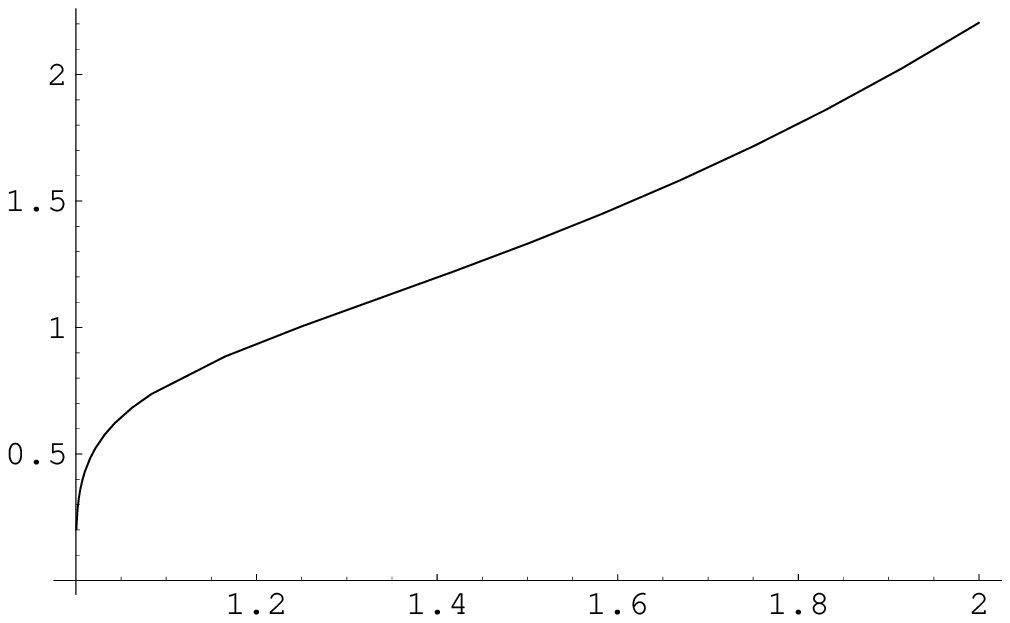}~~~
\includegraphics[height=2in]{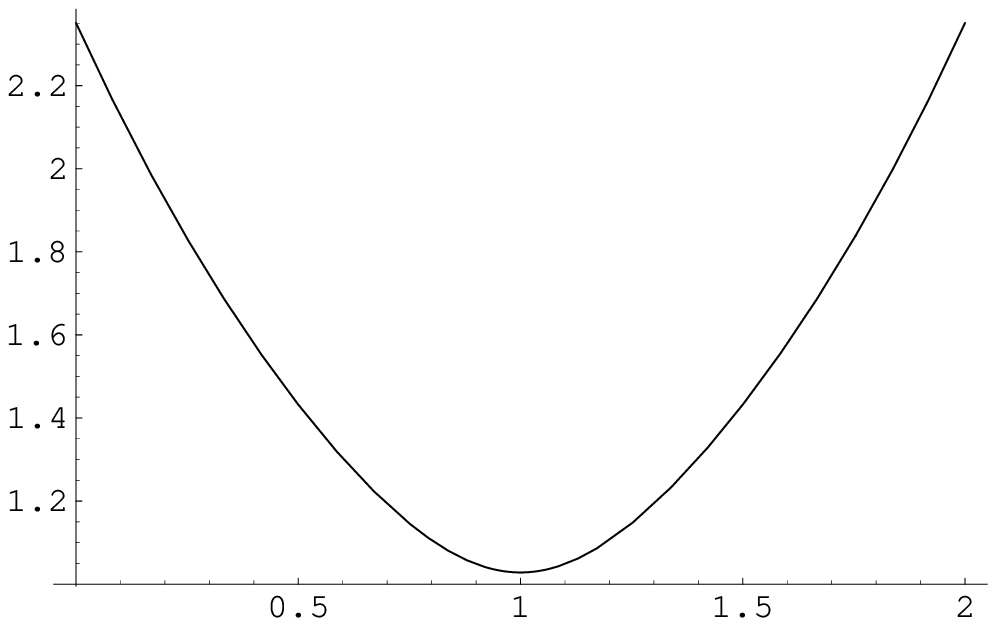}\\
\vspace{1mm}
Fig.1(a)~~~~~~~~~~~~~~~~~~~~~~~~~~~~~~~~~~~~~~~~~~~~~~~~Fig.1(b)\\

\vspace{5mm} Fig. 1(a) shows the variation of the scale factor
(given by eq.(42)) over the time for
$\kappa=2,~B=\rho_{0}=3,~C=1$ and $t_{0}=1$ while Fig. 1(b) shows
the same (for eq.(43)) for $\kappa=2,~B=\rho_{0}=3,~C=3$ and
$t_{0}=1$ In both the cases, (+)ve sign is taken. \hspace{1cm}
\vspace{6mm}
\end{figure}

{\bf Case-II:~~Bulk with Cosmological Constant}\\

In this case we choose in the bulk $p_{B}=p_{5}=-\rho_{B}$. It
follows then from Bianchi identity, i.e, the conservation
relation (21) demands $\dot{\rho_{B}}=0$, which in turn in view
of (20) yields $\rho_{B}'=0$. Hence $\rho_{B}$ is purely a
constant said to be the cosmological constant in five dimension
($\equiv\Lambda_{5}$). Now the generalized Friedmann like
equations (38) and (40) are modified in this case as

\begin{equation}
\frac{\dot{a_{0}}^{2}}{a_{0}^{2}}=\frac{\kappa^{2}}{6}\Lambda_{5}
+\frac{\kappa^{4}}{36}~\rho_{b}^{2}+\frac{C}{a_{0}^{4}}
\end{equation}

and

\begin{equation}
\ddot{a_{0}}=-\frac{\kappa^{4}}{36}\frac{(3A+2)}{(A+1)}
\frac{\rho_{0}}{a_{0}^{(6A+5)}}+\frac{\kappa^{4}Ba_{0}}{36(1+A)}
-\frac{\kappa^{2}}{6}\Lambda_{5}a_{0}-\frac{C}{a_{0}^{3}}
\end{equation}

We choose here $A=1/3$, so that at the~begining, the~brane world
may be said to be filled up with ($p=\rho/3$) radiation.\\

We now examine the following cases:\\

{\bf a)}~~When $\frac{\kappa^{4}B}{48}
-\frac{\kappa^{2}}{6}|\Lambda_{5}|<0$\\

For this restriction $\ddot{a_{0}}<0$ for all time while
$\dot{a_{0}}$ vanishes at some finite $a_{0}$. It is therefore, a
decelerating model of the universe with only a maximum but no
minimum. It is a recollapsing~brane model (see fig. 2(a)) with
$a_{0}\rightarrow0$ and $\ddot{a_{0}}\rightarrow-\infty$ at both
ends (radiation to radiation).\\

{\bf b)}~~When $\frac{\kappa^{4}B}{48}
-\frac{\kappa^{2}}{6}|\Lambda_{5}|>0$\\

Here $H_{0}$ is non-zero throughout the evolution. As
$a_{0}\rightarrow0,~\ddot{a_{0}}\rightarrow-\infty$ and as
$a_{0}\rightarrow\infty,~\ddot{a_{0}}\rightarrow\infty$. So there
is a transition from~deceleration to acceleration in course of
evolution of the~brane (Graphically it is shown in fig. 2(b)).\\

{\bf c)}~~When $\frac{\kappa^{4}B}{48}
-\frac{\kappa^{2}}{6}|\Lambda_{5}|=0$~~~~i.e,~~~~
$|\Lambda_{5}|=\frac{\kappa^{2}B}{8}$\\

In this case there is always $\ddot{a_{0}}<0$ with
$H_{0}^{2}\rightarrow0$ as $a_{0}\rightarrow\infty$. So this has
a typical Einstein-de Sitter like behaviour for the spatially
flat standard model (presented in fig. 2(c)). The interesting
feature of such a~brane world model is that even though at large
$a$ the fluid has an equation of state $p_{b}=-\rho_{b}$, the
model does not show the character of a $\Lambda$CDM model as in
standard cosmology. Rather it still remains a decelerating model.
As we note that in one of the above three models of the~brane
universe, for example in the second model, there occurs a
transition from a decelerating to an accelerating phase. It may
be worthwhile to locate the said transition point where
$\ddot{a_{0}}=0$. It is, however, not difficult to find out from
the equation (46) that the~brane world model with $A=1/3$ remains
accelerating
($\ddot{a_{0}}>0$) as long as\\

~~~~~~~~~~~~~~~~~~~~~~~~~$-\frac{\kappa^{4}}{16}\frac{\rho_{0}}{a_{0}^{7}}
+\left(\frac{\kappa^{4}}{48}B-\frac{\kappa^{2}}{6}|\Lambda_{5}|\right)a_{0}
-\frac{C}{a_{0}^{3}}>0$\\

that is,\\

~~~~~~~~~~~~~~~~~~~~~~~~~~~~~~~~$a_{0}^{8}>\frac{\frac{\kappa^{4}}{16}\rho_{0}
+Ca_{0}^{4}}{\frac{\kappa^{4}}{48}B-\frac{\kappa^{2}}{6}|\Lambda_{5}|}$,\\

which finally reduces to the following condition by a
straightforward calculation:\\

~~~~~~~~~~~~~~~~~~~~~~~~~$a_{0}^{4}>\frac{C}{2\left(\frac{\kappa^{4}}{48}B
-\frac{\kappa^{2}}{6}|\Lambda_{5}|\right)}\left[1
+\sqrt{\frac{\kappa^{4}}{4}\frac{\rho_{0}}{C^{2}}\left(\frac{\kappa^{4}}{48}B
-\frac{\kappa^{2}}{6}|\Lambda_{5}|\right)}\right]$\\

In the standard cosmology with the~chaplygin gas in FRW universe,
the corresponding inequality is\\

~~~~~~~~~~~~~~~~~~~~~~~~~~~~~~~~~~~~~~~~~~~~~~~$a_{0}^{8}>\rho_{0}/B$.\\

Note that, this is only a formal comparison, as $\rho_{0},~B$ etc.
in the~brane and the corresponding quantities in FRW universe are
not really identical.\\

\begin{figure}
\includegraphics[height=2in]{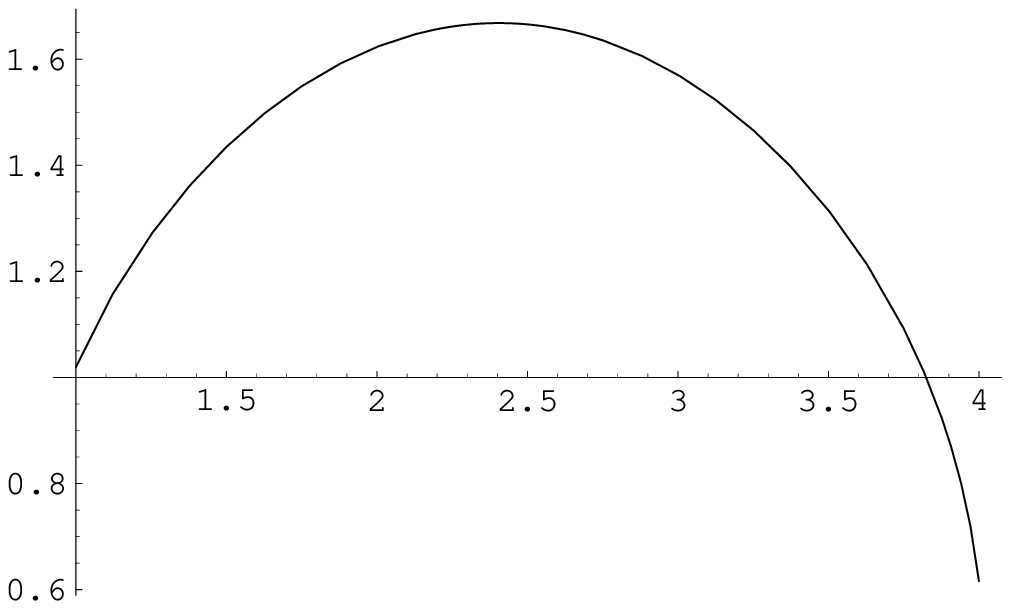}~~~
\includegraphics[height=2in]{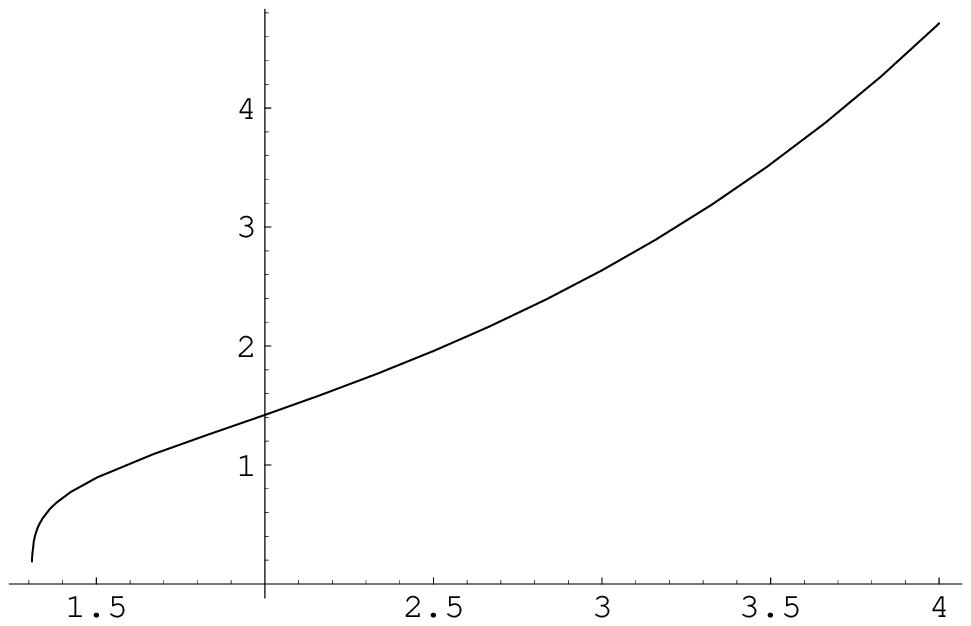}\\
\vspace{1mm}
Fig.2(a)~~~~~~~~~~~~~~~~~~~~~~~~~~~~~~~~~~~~~~~~~~~~~~~~Fig.2(b)\\
\vspace{5mm} \hspace{1cm} \vspace{6mm}

\includegraphics[height=2in]{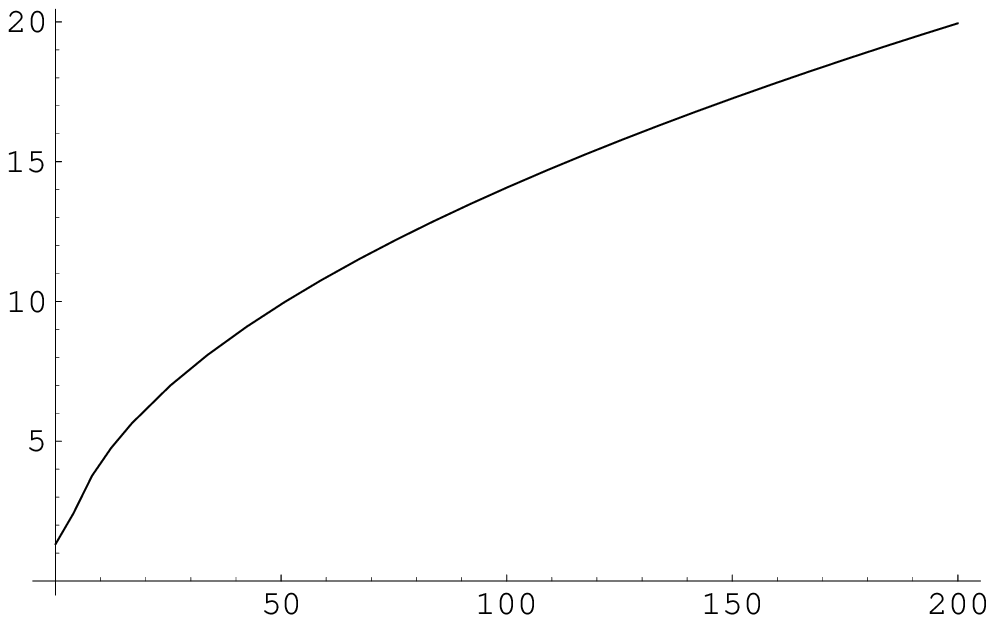}
\vspace{1mm}
Fig.2(c)~~~~~~~~~~~~~~~~~~~~~~~~~~~~~~~~~~~~~~~~~~~~~~~~~~~~~~\\

\vspace{5mm} Fig. 2(a) shows the~recollapsing model of the
universe for $\kappa=1,~B=1,~\rho_{0}=48,~C=1,~|\Lambda_{5}|=1$
and $t_{0}=24$.\\
Fig. 2(b) shows a transition from~deceleration to acceleration for
$\kappa=1,~B=16,~\rho_{0}=24,~C=\frac{1}{2},~|\Lambda_{5}|=1$ and
$t_{0}=1$. Lastly, Fig. 2(c) shows the Einstein-de Sitter
universe for $\kappa=1,~B=8,~\rho_{0}=48,~C=1,~|\Lambda_{5}|=1$
and $t_{0}=1$. \hspace{1cm} \vspace{6mm}
\end{figure}

\section{\normalsize\bf{Concluding remarks}}

When the bulk is filled with only dust, there are certain
restrictions imposed on the metric as well as on the matter
content of the~brane. As results of such restrictions the metric
coefficient $g_{tt}$ becomes independent of the y-coordinate,
while the perfect fluid in the~brane satisfies an effective
equation of state $p_{B}=-\frac{2}{3}\rho_{B}$, which apparently
indicates that matter in the~brane is in the form of dark
energy.\\

On the other hand, if the matter in the bulk is in the form of
cosmological constant or vacuum, there is neither any restriction
on the metric nor on the matter contained by the~brane.
Bin\'{e}truy et al and Leon [3,~4] have shown perfect fluid
solution with ~barotropic equation of state for~brane matter. In
this paper, we have shown that for~brane matter, it is possible
to have ~chaplygin gas form of fluid and obtained some interesting
conclusions. The~recollapsing model of the~brane universe shows
that in course of evolution it reaches a maximum size and finally
collapses to a singularity. So for such a model, there is
~deceleration through the entire evolution period and it has
singularities at both ends. This difference in behaviour of the
model is caused by the modifications introduced in the field
equations by the presence of an anti-de Sitter bulk
($\Lambda_{5}<0$). Another point of significance for a~chaplygin
gas in the~brane world model is that, the density does not vanish
as the scale factor increases indefinitely, rather it reaches a
finite magnitude with an equation of state $p_{b}=-\rho_{b}$.
This behaviour still remains valid in one of the~brane world
models, which expands indefinitely with~deceleration throughout
and resembles the standard Einstein-de Sitter model. Such
behaviour of the~brane is in clear contradiction to the standard
$4D$ Friedmann universe.\\

{\bf References}:\\
\\
$[1]$ L.Randall and R.Sundrum, {\it Phys.Rev.Lett} {\bf 83}, 3370
(1999); {\it Phys.Rev.Lett} {\bf 83}, 4690 (1999).\\
$[2]$ T.Shiromizu, K.Maeda and M.Sasaki, {\it Phys.Rev.D} {\bf
62} 024012 (2000).\\
$[3]$ P.Bin\'{e}truy, C.Deffayet, U.Ellwanger and D.Langlois,
{\it Phys.Letts.B} {\bf 477}, 285 (2000); P.Bin\'{e}truy,
C.Deffayet, U.Ellwanger and D.Langlois, {\it Nucl.Phys.B} {\bf
565}, 269 (2000).\\
$[4]$ J.Ponce de Leon, {\it Mod.Phys.Letts.A} {\bf 17}, 2425
(2002); {\it Mod.Phys.Letts.A} {\bf 16}, 2291 (2001).\\
$[5]$ V.Gorini, A.Kamenshchik and U.Moschella, {\it Phys.Rev.D}
{\bf 67}, 063509 (2003).\\
$[6]$ S.J.Perlmutter et al, {\it Astrophys.J.} {\bf 517}, 565
(1999).\\
$[7]$ P.J.E.Peebles and B.Ratra, {\it Rev.Mod.Phys.} {\bf 75}, 559
(2003); T.Padmanabhan, {\it Phys.Rept.} {\bf 380}, 235 (2003);
A.Krause and Siew-Phang Ng, {\it Int.J.Mod.Phys.A} {\bf
21}, 1091 (2006) and references therein.\\
$[8]$ J.Ponce de Leon, {\it gr-qc}/0511150.\\
$[9]$ J.Ponce de Leon, {\it Gen.Relt.Grav.} {\bf 37}, 53 (2005).\\
$[10]$ J.Ponce de Leon, {\it Class.Quant.Grav.} {\bf 20}, 5321
(2003).\\
$[11]$ E.Papantonopoulos, {\it hep-th}/0202044.\\
$[12]$ U.Debnath, A.Banerjee and S.Chakraborty, {\it
Class.Quant.Grav.} {\bf 21}, 5609 (2004).\\

\end{document}